# Characterizing Dislocation Substructures in Creep-Deformed Olivine Using Electron Channeling Contrast Imaging

M. Haroon Qaiser[1], Jessica White[2*], David Wallis[2], and T. Ben Britton[1]

[1]Department of Materials Engineering, University of British Columbia, Vancouver, BC, Canada

[2]Department of Earth Sciences, University of Cambridge, Cambridge, United Kingdom

[*]Present address: Department of Earth Sciences, University of Southern California, Los Angeles, CA, United States

Correspondence to: Ben Britton, ben.britton@ubc.ca

## Graphical Abstract

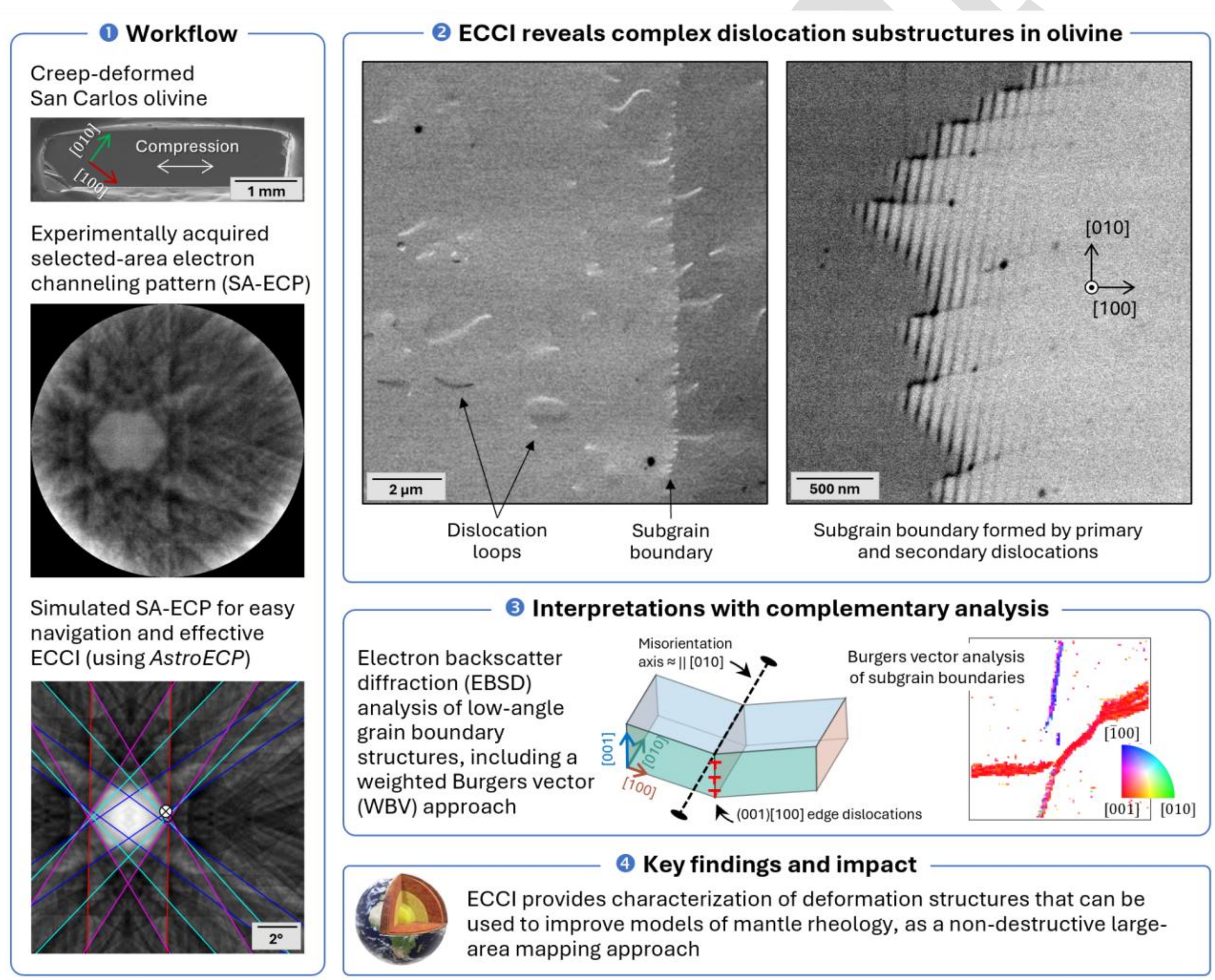

## Key Points

- Electron channeling contrast imaging on bulk olivine reveals subgrain boundaries and dislocations without using TEM or decoration techniques
- Despite the limited slip systems, subgrain boundaries can be complex in olivine, which can be helpful for informing rock deformation models
- Weighted Burgers vector mapping from electron backscatter diffraction confirms multiple dislocation types within subgrain boundaries

## Abstract

Olivine is the dominant mineral in Earth's upper mantle and therefore controls mantle rheology and the mechanics of plate tectonics. The constitutive laws for dislocation-mediated deformation of olivine depend on the nature, density, and arrangements of dislocations within crystals. Hence, imaging and characterizing these defects is important, albeit challenging. Traditional imaging approaches involve (1) transmission electron microscopy (TEM), which samples small areas and requires extensive preparation and (2) oxidation decoration methods that have low spatial resolution and cannot distinguish dislocations of opposite Burgers vectors. Here, we apply electron channeling contrast imaging (ECCI) to unlock insight into the deformation structures within olivine, and combined with electron backscatter diffraction (EBSD) and weighted Burgers vector (WBV) mapping as an informative route to characterize dislocation substructures in bulk materials. Specifically, we have used an ECCI workflow based on selected-area electron channeling patterns (SA-ECPs) and we apply this workflow to a single crystal of San Carlos olivine that was deformed by creep at high temperature. ECCI micrographs reveal subgrain boundaries, surface threading dislocations, and dislocation loops across representative areas. The observations demonstrate that this workflow can reliably reveal the complexity of subgrain boundaries in olivine, which can host multiple dislocation types and exhibit non-planar geometries. Despite the limited number of slip systems in olivine, subgrain boundaries can form complex, mixed assemblies. Overall, such observations can provide a variety of constraints on dislocation types, morphologies, and distributions, which are required to parameterize and calibrate models of transient and steady-state dislocation creep in olivine and other materials.

## Plain Language Summary

Earth's mantle lies beneath the crust and makes up an important part of the planet's interior as it controls many tectonic processes. Olivine is the most abundant mineral in the upper mantle and therefore strongly influences how mantle rocks deform and flow. Rocks in the mantle deform because their crystals contain tiny defects called dislocations, and the number and arrangement of those defects controls how easily rocks can deform. Direct observation and imaging of these dislocations can be helpful in developing or validating deformation models. Traditional methods to image dislocations either sample extremely small areas or require destructive chemical treatments that can obscure important details. In the present work, we use a non-destructive electron imaging method called electron channeling contrast imaging, together with measurements of crystal orientations, to image and map dislocations across representative areas of olivine. This approach reveals a wide variety of defect structures and gives a richer picture of how mantle minerals accommodate strain, which can improve models of rock strength and deformation.

# 1. Introduction

Olivine is the most abundant mineral in Earth's upper mantle and therefore controls its rheological behavior and large-scale geodynamics (Avé Lallemant & Carter, 1970). Therefore, constitutive laws for creep of olivine have been established via laboratory deformation experiments at high temperature, which are used to predict flow behaviors under upper-mantle conditions (e.g., Bai et al., 1991; Hansen et al., 2011; Keefner et al., 2011). Such flow laws predict that dislocation-mediated deformation dominates much of the shallow upper mantle (e.g., Warren & Hansen, 2023). Recent works have attempted to improve the reliability of such extrapolations by expounding the physical basis of the constitutive equations in terms of interactions among dislocations, allowing the development of formulations that describe both transient and steady-state creep (Breithaupt et al., 2023; Hansen et al., 2019, 2021; Holtzman et al., 2018; Mulyukova & Bercovici, 2022). These interactions depend on the dislocation density, their arrangements and spatial distribution, and the partitioning of dislocations among different types and between different signs of their Burgers vectors (e.g., Wallis et al., 2017, 2020, 2021; Wiesman et al., 2024). However, the techniques most commonly used to resolve individual dislocations, which are transmission electron microscopy (TEM; e.g., Mussi et al., 2017; Phakey et al., 1972) and oxidation decoration (Karato, 1987; Kohlstedt et al., 1976), have some important limitations. TEM delivers exceptional spatial resolution but samples only very small volumes that are accessed using complicated sample preparation. Decoration methods are locally destructive in nature, have limited spatial resolution, and do not reveal the sign of the Burgers vector.

Electron channeling contrast imaging (ECCI) is a scanning electron microscopy (SEM) technique that has potential to address the above challenges. In ECCI, electrons channel with regard to the atomic structure and crystallographic lattice planes in the sample. Contrast depends on the ease or difficulty of 'channeling-in'. This effect modulates the backscatter intensity and thereby generates contrast that can be sampled effectively, for example, using a backscattered electron (BSE) detector. Many researchers have explored and presented the principles, and the reader is referred to these studies for further detail (e.g., Crimp, 2006; Lloyd, 1987; Picard et al., 2014; Qaiser et al., 2026; Wilkinson & Hirsch, 1997; Zaefferer & Elhami, 2014). As examples, in the case where a sample contains multiple grain orientations and the atomic-number contrast is relatively low, individual grains will have different channeling effects based on their crystal orientation, which enables grains to appear as regions of common gray level. In the case where a sample contains dislocations and other crystallographic defects, the contrast is based on the sensitivity of electron channeling to local lattice disturbances and the elastic strain field around a dislocation. ECCI has been applied in many case studies on a variety of materials to effectively image dislocations, subgrain boundaries, and other crystallographic defects (e.g., Mansour et al., 2015; Miyajima et al., 2018; Weidner & Biermann, 2015). In particular, Miyajima et al. (2018, 2019) used a so-called orientation-optimized ECCI method to image dislocations in olivine without using selected-area electron channeling patterns (SA-ECPs) or the controlled ECCI technique (Zaefferer &

Elhami, 2014). However, ECCI as an approach has potential to be refined by direct use of the indexed ECP to select the channeling condition and optimize contrast within the material, which we explore in this contribution.

Recently, we have developed a new ECCI workflow based on using SA-ECPs to determine the local crystal orientation, navigate the accessible channeling conditions, and achieve the precise alignment of the specimen and beam required to optimize channeling contrast with the finesse required to characterize crystal defects (Qaiser et al., 2026). This approach has advantages over other variants of ECCI that use electron backscatter diffraction (EBSD) to determine the geometry required for channeling contrast (Miyajima et al., 2018; Zaefferer & Elhami, 2014). For example, the crystallographic analysis of the SA-ECP is conducted in broadly the same geometry as the ECCI and thereby circumvents the need to switch between geometries and introduce associated inaccuracy and/or imprecision. This improved precision is enhanced further by the improved resolution of the Kikuchi pattern that SA-ECPs offer relative to electron backscatter patterns (EBSPs), which enables more precise control over the crystal-beam geometry and associated channeling contrast. Therefore, this paper aims to provide a demonstration of this approach in the context of characterizing dislocations and subgrain boundaries in deformed olivine to highlight key points in the application of the technique and demonstrate the quality of data that can be gained. We combine the SA-ECP informed ECCI with complementary EBSD data to present an exposition of the combined use of multiple spatially correlated techniques for dislocation analysis to demonstrate levels of detail that can be achieved in the SEM.

## 2. Materials and Methods

We investigate a natural single-crystal sample of olivine, approximately $Fo_{90}$, from San Carlos, Arizona (Figure 1). This sample was deformed in a set of experiments for which some of the steady-state creep data were published by Ricoult and Kohlstedt (1985). However, this particular sample (sample number 83OL-EN3(1), experiment number 8306) was not featured in that paper. The sample was previously compressed at 45° to [100] and [010] (i.e., the $[110]_c$ orientation in the notation of Durham and Goetze, 1977) at a temperature of 1300°C and a constant stress of ~50 MPa. The sample orientation and compression direction are labeled in Figure 1a. The axis conventions used in this study are aligned with the *Pbnm* setting of space group 62, with <*a*> as the shortest axis and <*b*> as the longest, consistent with many of the previously reported studies on olivine (e.g., Demouchy, 2021; Mussi et al., 2017). The reader is therefore advised caution when comparing literature and data from different software, which might be using *Pnma* settings, as that changes the axis convention described in this characterization.

The sample was mechanically polished, with a final fine polishing step down to 0.05 μm using a colloidal silica suspension. This mechanical polishing was followed by broad ion beam (BIB) polishing at 8 keV and application of a thin carbon coat using a Gatan Model 685 Precision Etching and Coating System (PECS II, Gatan Inc. USA). The sample was prepared on the (001) plane to be able to see dislocations on the (010)[100] slip system, which is the slip system expected to be active based on the compression direction and is the weakest slip system in dry olivine at high temperatures (Bai et al., 1991; Demouchy, 2021; Durham & Goetze, 1977; Durham et al., 1977). However, we note that preparing a (010) surface could also be useful to further distinguish between dislocation types with [100] and [001] Burgers vectors. The total density of free dislocations in the sample is predicted to be approximately $4\times10^{11}$ $m^{-2}$ based on the piezometer of Bai and Kohlstedt (1992).

The ECCI and EBSD experiments were carried out in a TESCAN AMBER-X plasma focused ion beam field-emission scanning electron microscope equipped with a retractable four-quadrant backscattered electron (4Q-BSE) detector below the pole piece and a Symmetry S2 EBSD detector. Beam energies of 20 keV to 30 keV and beam currents of 1 nA to 10 nA were used in this study. For ECCI in the SEM, initially the sample was placed so that the electron beam was at normal incidence to the sample, and then small stage tilts (± 5°) were applied to navigate to specified channeling conditions indicated by the SA-ECPs. The tilts and navigation to a channeling condition were informed by AstroECP, with the detailed method reported elsewhere (Qaiser et al., 2026). EBSD was carried out in a 70° tilted configuration using a pre-tilted holder. Three EBSD datasets were acquired—a full-sample dataset, subgrain-boundary data, and local data for weighted Burgers vector (WBV) analysis, with step sizes of 8 μm, 150 nm, and 120 nm, respectively. The initial orientation data acquired via the Hough transform were refined by pattern matching with dynamical simulated patterns (Winkelmann et al., 2021) using MapSweeper in AZtecCrystal v3.3 (Oxford Instruments Inc. UK). EBSD data analysis and plotting was conducted using the MTEX v5.11.2 toolbox for MATLAB (Bachmann et al., 2010).

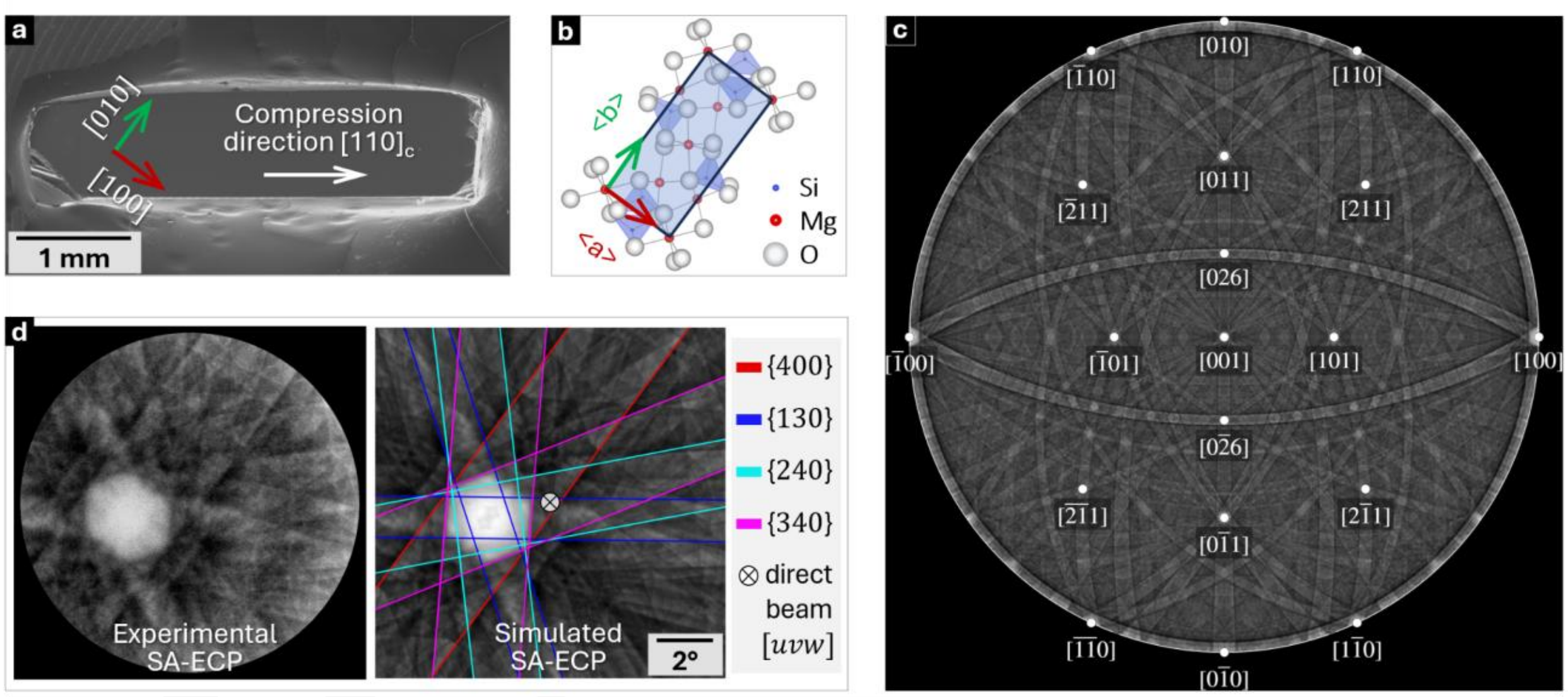

**Figure 1. (a)** Secondary electron image of the sample. **(b)** Crystal structure of forsterite along with the conventions used for lattice constants in an orthorhombic crystal. **(c)** Dynamically simulated stereogram of olivine with labeled zone axes, simulated in EMsoft at a primary beam energy of 20 keV. **(d)** Experimental SA-ECPs of the sample in the orientation shown in (a) and (b) along with its simulated and indexed pattern, reprojected from the stereogram using AstroECP.

ECCI was carried out by tilting and rotating the sample to target specific channeling conditions, aligning the direct beam direction [*uvw*] along the edges of high-contrast Kikuchi bands. Post-processing of ECPs or BSE micrographs, where indicated, includes a fast Fourier transform (FFT) based bandpass filter designed to remove background low-frequency, long-range gradients. The raw data and images are shared in a Zenodo repository (see the data availability statement).

For dynamical simulation of forsterite, we used EMsoft (Singh et al., 2017) to generate high-resolution 2000 × 2000 pixel Kikuchi stereograms at 20 keV and 30 keV (e.g., Figure 1c). Each stereographic projection was stored to disk, and ECPs were generated by reprojecting the stereogram using the open-source graphical user interface AstroECP (Qaiser et al., 2026), which is based on the approaches used in AstroEBSD (Britton et al., 2018) and MTEX (Bachmann et al., 2010). Figure 1d shows a representative experimental SA-ECP from the sample as well as its simulated and indexed reprojection from AstroECP.

WBV analysis was carried out in MATLAB using pattern-matching-based EBSD data and an MTEX built-in function (`weightedBurgersVec`) based on Wheeler et al. (2009) and Wheeler et al. (2024). For the analysis here, the path integral method was used and the window size was 1 (i.e., data were calculated from a square region of 9 pixels), and a lower-bound threshold of 0.004 $\mu m^{-1}$ was applied to exclude noise. The WBV direction was transformed into crystal coordinates and colored according to the inverse pole figure (IPF) (red = [001], green = [010], blue = $[\bar{1}00]$).

# 3. Results

## 3.1. Large-Area EBSD Analysis

A large-area EBSD map was acquired to ascertain the sample orientation after deformation. Figure 2 presents IPF maps of the entire sample, indicating that the sample remains a single crystal close to the $[110]_c$ orientation. The subtle color gradient in the IPF maps is due to a long-range rotation about the [001] axis, as indicated by the lack of dispersion of the Z-axis, which is subparallel to [001] in the IPF-Z plot. The kernel average misorientation (KAM) map highlights the presence of subgrain boundaries. To enhance the contrast of very low angle grain boundaries, the color scale in Figure 2b is capped at a maximum value of 1° but regions within the map have KAM values of up to 3°.

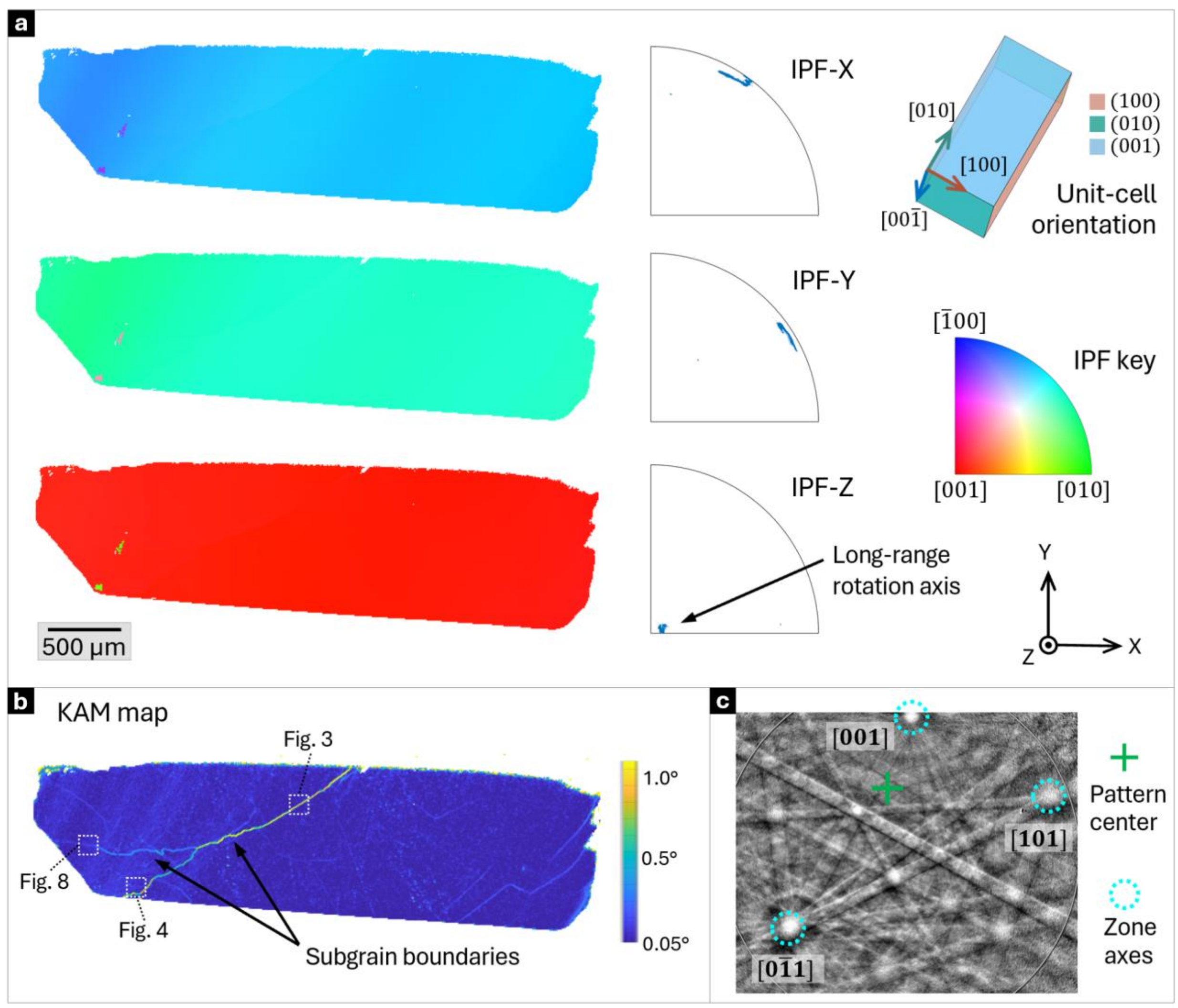

**Figure 2. (a)** IPF maps of the single-crystal olivine sample. **(b)** Kernel average misorientation (KAM) map indicating the presence of subgrain boundaries. Regions of interest for further analysis are marked (not to scale). **(c)** A representative EBSP highlighting the relevant (low-index) zone axes.

To further inspect the most prominent subgrain boundary shown in Figure 2b, we used more detailed EBSD analysis. In Figure 3a, boundaries are colored by the angle between their trace and the misorientation axis, so the bottom of the color scale indicates a small angle and therefore likely tilt character. Boundary segments with large angles, only a few of which are present, are ambiguous as they can either be twist or mixed, as discussed below. Figure 3b presents a sketch of the inferred tilt-boundary plane, the rotation axis, and the dominant edge character of the constituent (001)[100] edge dislocations. The misorientation diagrams in Figure 3c present the same misorientation axes in both sample and crystal coordinates, which highlights the alignment of the rotation axis with the boundary trace, consistent with an edge-dominated tilt boundary. The same area was further analyzed in a flat configuration, and SA-ECPs were acquired from either side of the boundary. The experimental and simulated SA-ECPs in Figure 3d further corroborate the inferred rotation axis and thus support the tilt classification. In some cases, the rotation axis can be interpreted from a pair of diffraction patterns through identification of the invariant point (i.e. the rotation axis) which may be out of the gnomonic projection, however in general the rotation axis can be identified from full indexing of the pattern and subsequent matrix analysis to calculate the axis/angle representation of the misorientation between the two crystals. The total misorientation angle of the boundary was found to be 1.77° ± 0.05° and 1.79° ± 0.05° with SA-ECP and EBSD, respectively.

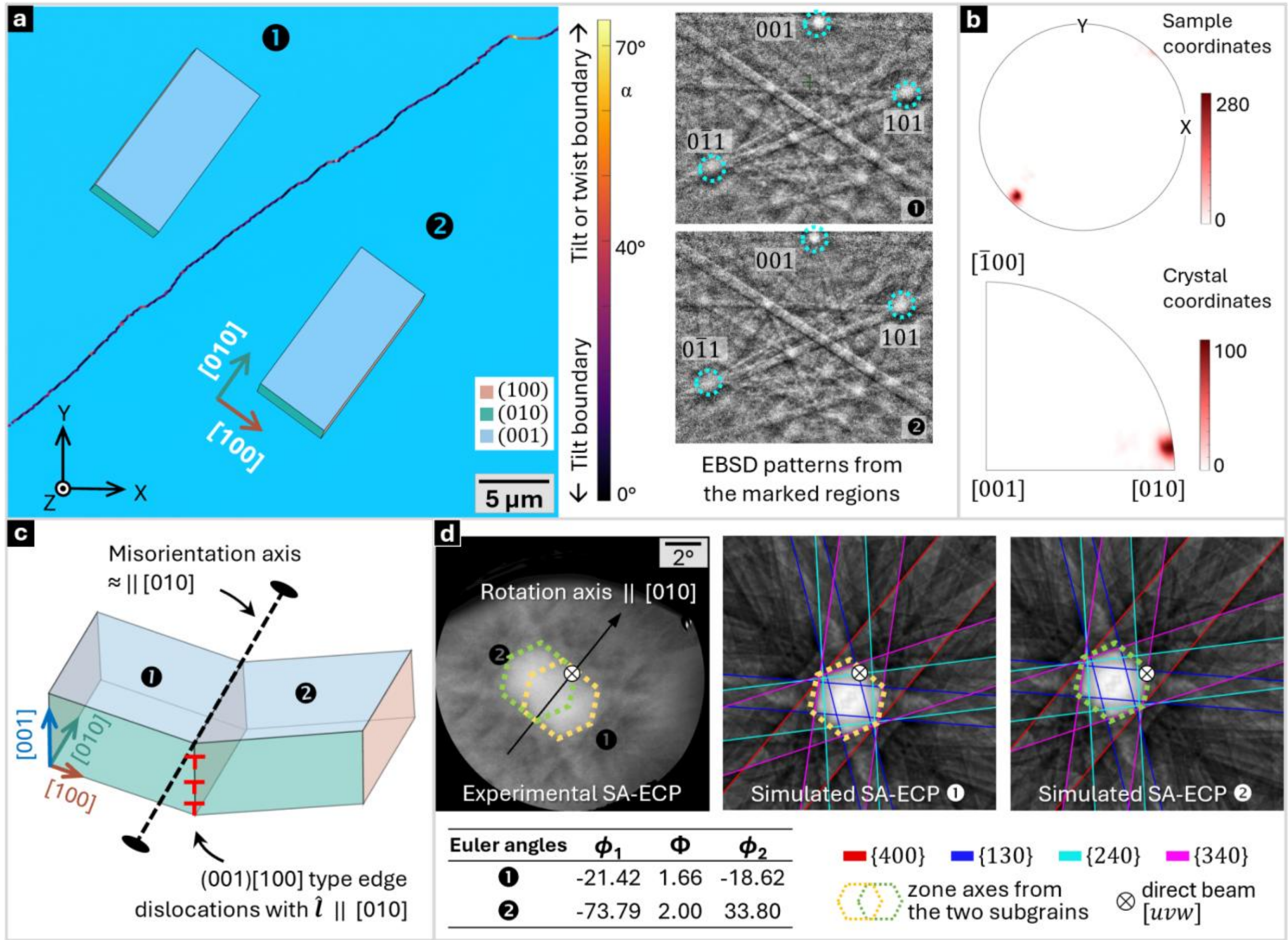

**Figure 3.** EBSD and SA-ECP analysis of a subgrain boundary. **(a)** IPF-X map of a selected region of the subgrain boundary, with unit cells overlaid and EBPS shown from either side of the boundary, i.e., the regions marked as 1 and 2. The color scale shows the angle between the boundary trace and misorientation axis. **(b)** Density distribution of misorientation axes between regions 1 and 2 depicted in sample coordinates and crystal coordinates. **(c)** Schematic of the subgrain boundary, indicating the tilt-boundary plane, misorientation axis, and type of dislocations forming the boundary. **(d)** Experimental and simulated SA-ECPs from regions 1 and 2, also indicating the same misorientation axis.

## 3.2. ECCI Analysis

Next, ECCI was carried out on the sample by using a channeling condition shown in Figure 4d. Note that this SA-ECP contains overlapping orientations from a relatively large selected area, ~500 μm across (see Section 4.1 for details). Subgrains were revealed in the ECCI micrographs as regions of common gray value, and the misorientations were measured through indexing their SA-ECPs. Surface-threading dislocations are evident as black or white dots or lines contrasting with the background. Dislocations that are oblique to the surface are present, identified by their comet-like shape, and are likely associated with inclined Burgers vectors (Crimp et al., 2001; Miyajima et al., 2018). In Figure 4b, an array of dislocations is present forming the subgrain boundary. The misorientation across this boundary is only about 0.12°, as measured through indexed SA-ECPs acquired on either side of the boundary. From the orientations of this sample and the dislocations, we expect the observed edge dislocations to have a Burgers vector of $\vec{b} = [100]$ and a line direction of $\hat{l} = [001]$. Subsurface dislocation loops and surface-penetrating half loops are also present, similar to previous reports of these structures within olivine (Mussi et al., 2015, 2017) and other materials (Picard et al., 2012).

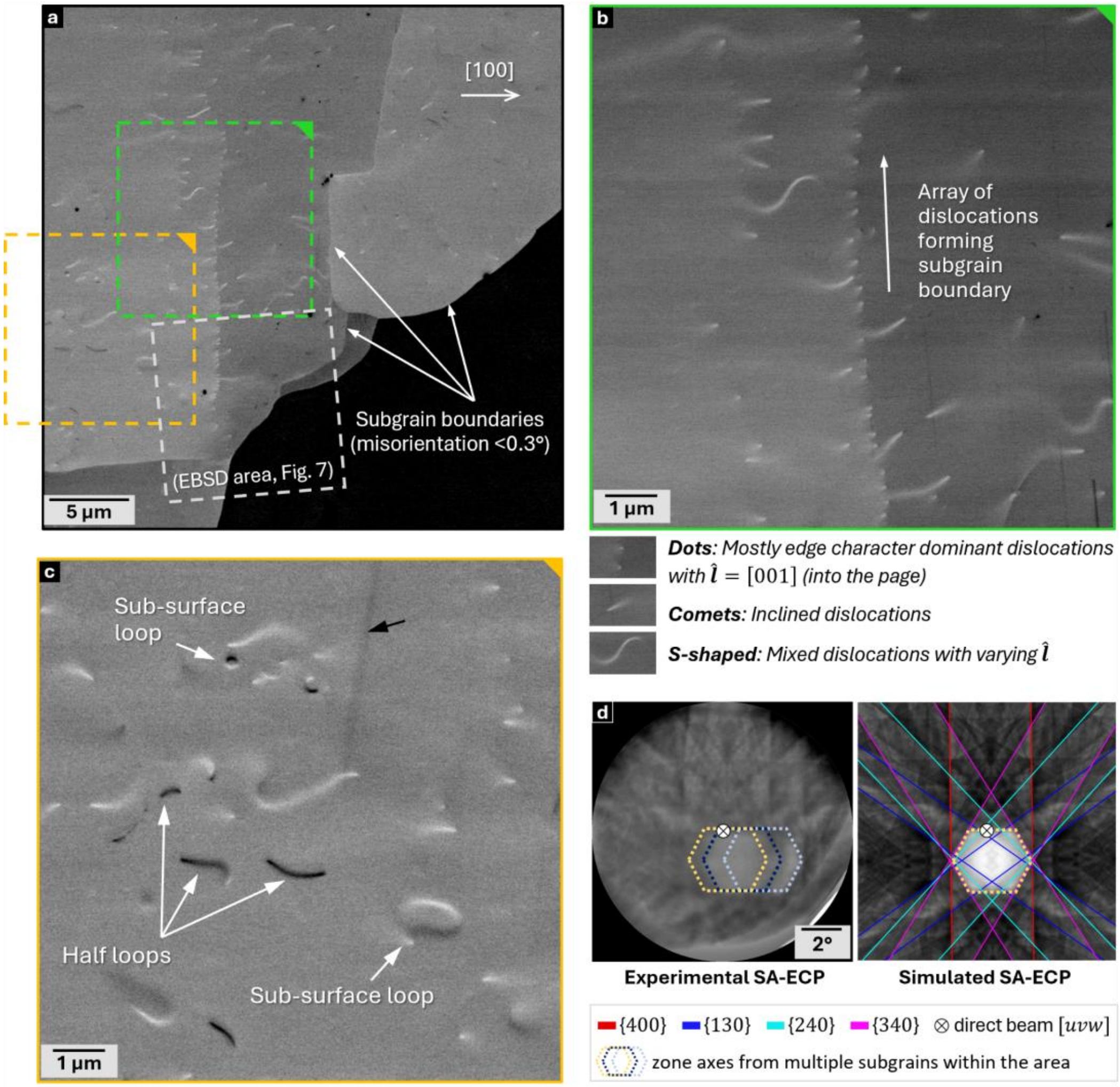

**Figure 4.** ECCI analysis of regions near subgrain boundaries. **(a)** ECCI micrograph of the region of interest illuminated under the channeling conditions shown by the SA-ECP in (d). A difference in grayscale intensity of subgrains can be seen as a result of local variations in electron channeling. **(b)** Higher-magnification image from the region marked in green in (a) showing an array of dislocations forming a subgrain boundary. Inclined and mixed-character dislocations can also be seen. **(c)** Higher-magnification image from the region marked in yellow in (a), featuring dislocation loops. The black arrow points to an artifact associated with carbon contamination from prior imaging. **(d)** Experimental and simulated electron channeling patterns from the region shown in (a).

The subgrain boundary from Figure 4 was characterized in more detail to analyze the dislocation structure and to quantify the misorientation across the boundary (Figure 5). We determined the misorientation angle, $\theta$, using three independent methods. First, we applied Read's formula, $\theta \approx \frac{|\vec{b}|}{d}$, which relates $\theta$ to the magnitude of the Burgers vector, $|\vec{b}|$, and the inter-dislocation spacing, $d$ (Read, 1953, p. 158). Second, we measured $\theta$ by indexing experimental and simulated SA-ECPs from either side of the boundary and using their corresponding Euler angles (Figure 5b). Third, we derived $\theta$ from the EBSD map acquired over the same region. Considering dislocations from the (010)[100] slip system, which is the most favorable system under our specific loading conditions and viewing direction (Bai & Kohlstedt, 1992), the magnitude of the Burgers vector is $|\vec{b}|$ = <*a*> = 0.475 nm, i.e., the shortest lattice parameter of the olivine unit cell. The average inter-dislocation spacing was measured using ImageJ 1.54i as 303 nm. The misorientation values from the SA-ECP analysis and EBSD data are in good agreement at 0.29° ± 0.05° and 0.26° ± 0.05°, respectively. However, the misorientation angle of 0.09° from Read's expression differs from the others (Figure 5a).

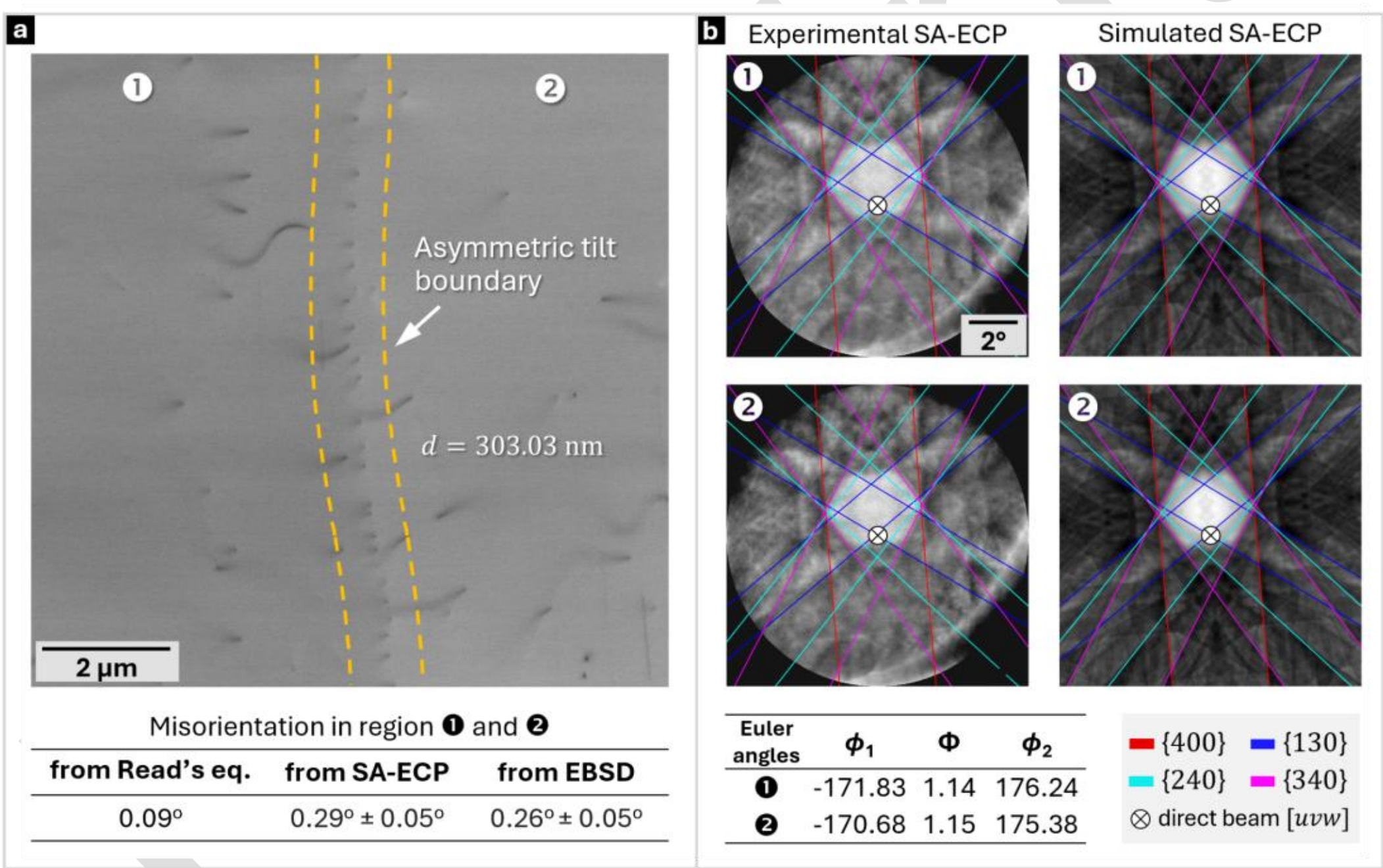

**Figure 5.** Further characterization of the subgrain boundary observed in Figure 4. **(a)** ECCI micrograph featuring the array of dislocations forming a subgrain boundary between the regions marked 1 and 2. The misorientation values, calculated using Read's expression, SA-ECP analysis, and EBSD, are also tabulated. **(b)** SA-ECPs taken from each side of the boundary (indexed using AstroECP) reveal subtle orientation differences, which were used to calculate the misorientation across the boundary.

To further explore the boundary and the possible cause of the deviation from Read's expression, we collected micrographs, presented in Figure 6, from the same area under five different channeling conditions. These ECCI micrographs reveal that the wall is not composed of a single, uniform dislocation type. At least four visually distinct dislocation types can be identified, which are annotated with different colored dots. For instance, if we compare the same regions from Figures 6d and 6b, the dislocations shown in the green, blue, and yellow boxes in Figure 6d all appear similar in contrast, however, the same set of dislocations are visually different from each other in Figure 6b. This analysis highlights an important aspect of the approach towards dislocation characterization through ECCI, in that we need to have micrographs taken at multiple channeling conditions to maximize the constraints on the character of dislocations. Additionally, in the channeling condition shown in Figure 6e, the contrast disappears for most of the dislocations, with only faint spots remaining, because the surface relaxation effects modify the local strain field and weaken the channeling contrast, especially for segments that are not optimally oriented to the electron beam (Picard et al., 2012; Picard et al., 2014; Wilkinson & Hirsch, 1995).

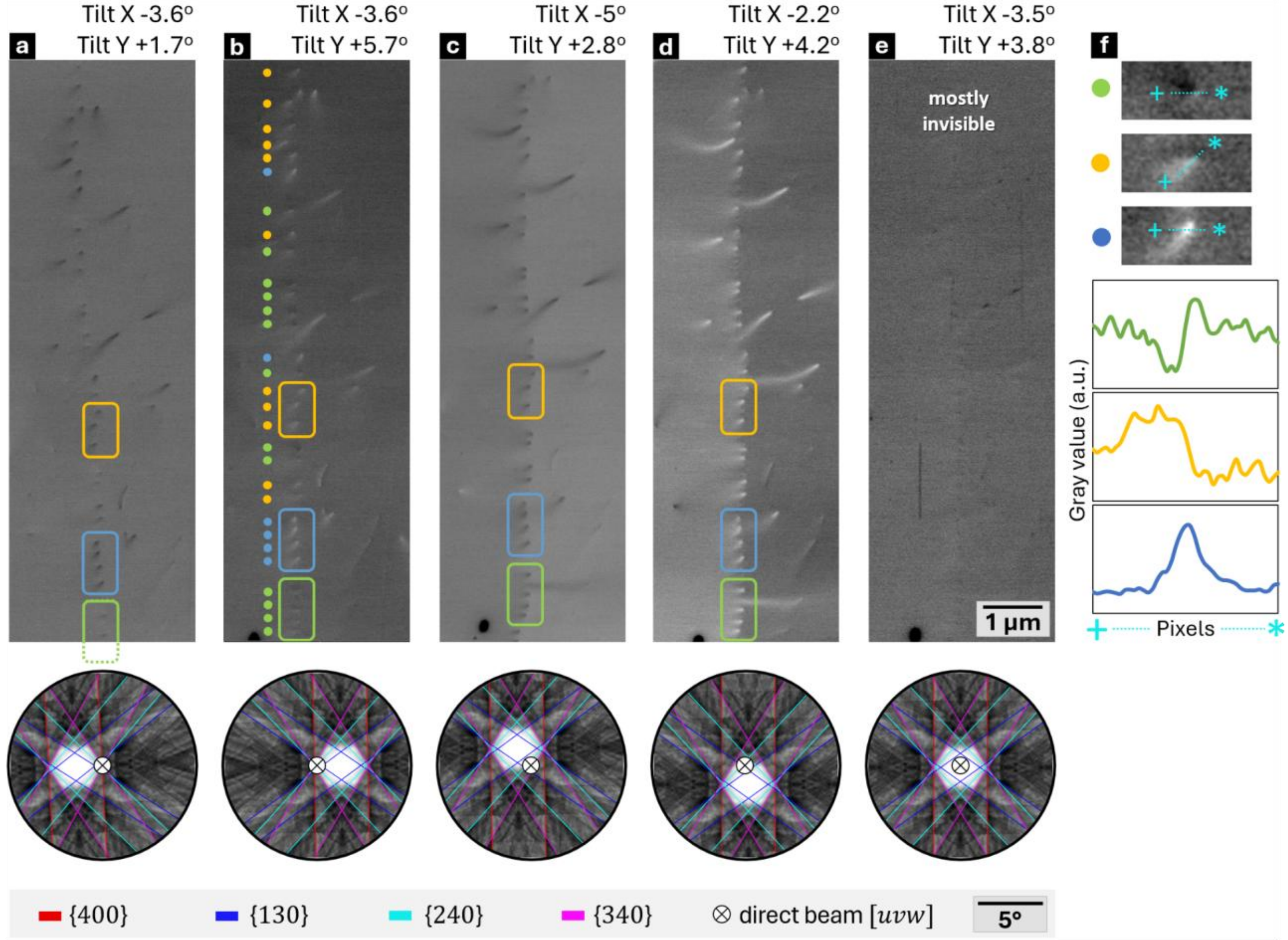

**Figure 6. (a–e)** The subgrain boundary from Figure 4b experimentally observed under different channeling conditions, which are shown by their corresponding SA-ECPs. The boundary appears to be composed of multiple types of dislocations, as denoted by the colored dots and boxes. **(f)** Enlarged examples of the three distinct dislocation types shown in (b), along with grayscale intensity profiles.

## 3.3. Weighted Burgers Vector Analysis of the Subgrain Boundary

Building on the ECCI observations from multiple channeling conditions, we performed WBV analysis in the region adjacent to the subgrain boundary to see the possible Burgers vectors associated with the dislocations forming the array. Figure 7a reproduces the ECCI micrograph region, featuring multiple subgrain boundaries and the dislocation array shown in Figure 4. The WBV analysis is shown in Figure 7b–d, where the direction map reveals two different types of boundary. One type is dominated by [001] Burgers vectors and the other is dominated by [100] Burgers vectors. The dislocation array of interest appears as a mixture of dark blue (i.e., $[\bar{1}00]$) and shades of pink (e.g., $[\bar{2}01]$, $[\bar{1}01]$, and $[\bar{2}03]$) in relation to the given IPF key. This indicates the presence of various weighted combinations of (010)[100] edge, (100)[001] edge, and [001] screw dislocations.

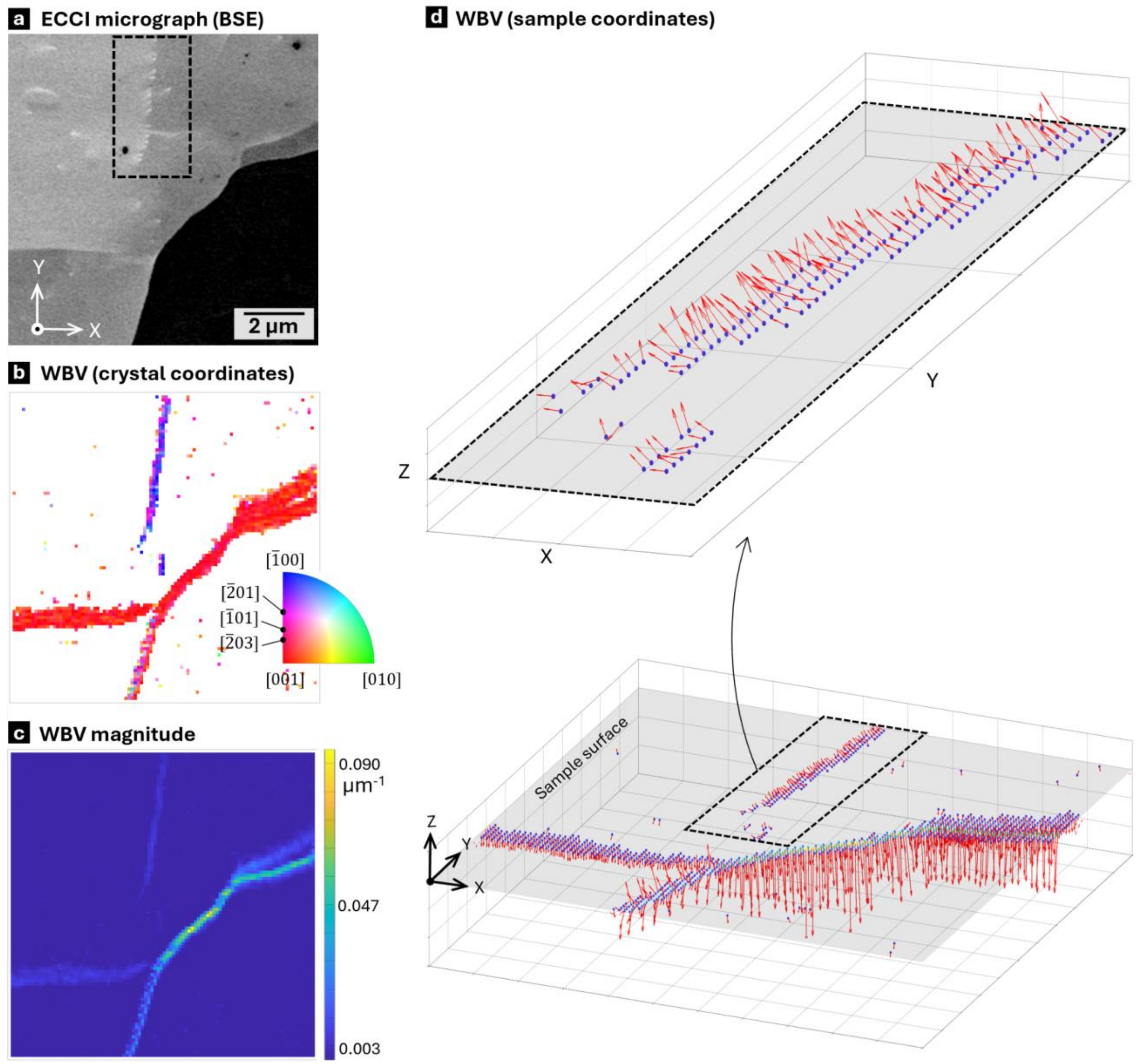

**Figure 7. (a)** ECCI micrograph in the flat-sample configuration and channeling condition as shown previously in Figure 4. The subgrain boundaries and dislocation array can be seen in the image contrast. **(b)** Directions of the WBV in crystal coordinates, colored by the IPF key shown in the inset. **(c)** Magnitude of the WBV, with a cutoff threshold of 0.004 $\mu m^{-1}$. **(d)** WBV representation in sample coordinates. The dots represent map data points colored by their WBV magnitude, with a cutoff threshold of 0.004 $\mu m^{-1}$. The arrows in red represent the measured WBV direction.

## 3.4. A Complex Subgrain Boundary

Next, the sample was rotated in the SEM to align its Y-axis with the [010] crystal direction, and a low-misorientation subgrain boundary region was imaged under the different channeling conditions indicated in Figures 8 and 9. The subgrain boundary has a ‘zig-zag’ pattern (Miyajima et al., 2018) and appears to be composed of primary and secondary dislocations. The line direction of evenly-spaced primary dislocations appears parallel to [010]. A similar-looking boundary wall was reported using TEM by Durham et al. (1977) in what was believed to be an apex of a tilt wall. Analysis of SA-ECPs taken from either side of the boundary shows that the rotation axis is parallel to the [010] crystal direction. As expected, grain contrast inverts between two opposite channeling conditions, marked as square vs. triangle symbols in the Figure 8. Some free dislocations are also visible, and the insets in Figures 8 and 9 show the directions of their black–white contrast. No classical pileups were observed in the sample under the given illumination conditions.

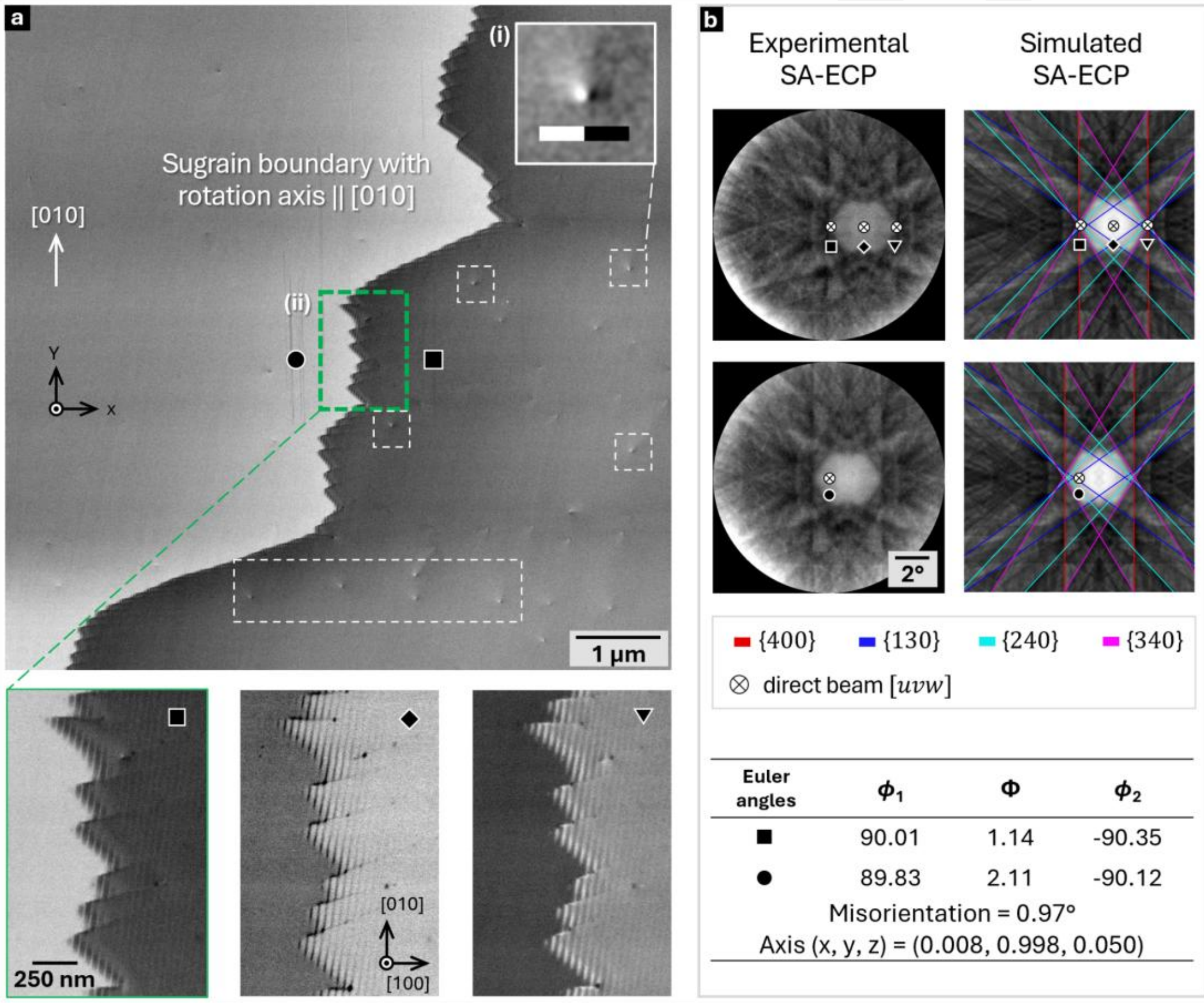

| Euler angles | $\phi_1$ | $\Phi$ | $\phi_2$ |
|---|---|---|---|
| ■ | 90.01 | 1.14 | -90.35 |
| ● | 89.83 | 2.11 | -90.12 |
| Misorientation = 0.97° | | | |
| Axis (x, y, z) = (0.008, 0.998, 0.050) | | | |

**Figure 8. (a)** ECCI micrograph featuring a subgrain boundary with an axis of rotation parallel to the [010] direction. Dashed white outlines indicate dislocations within the grain interior with $\hat{\boldsymbol{l}}$ = [001]. Inset (i) magnifies one of these dislocations to show the contrast around it. Inset (ii) shows the grain boundary at higher magnification, under different channeling conditions marked with geometric symbols. Each of these conditions are shown in (b). **(b)** Experimental and indexed/simulated SA-ECPs taken from either side of the boundary. The boundary characteristics are given in the table insert.

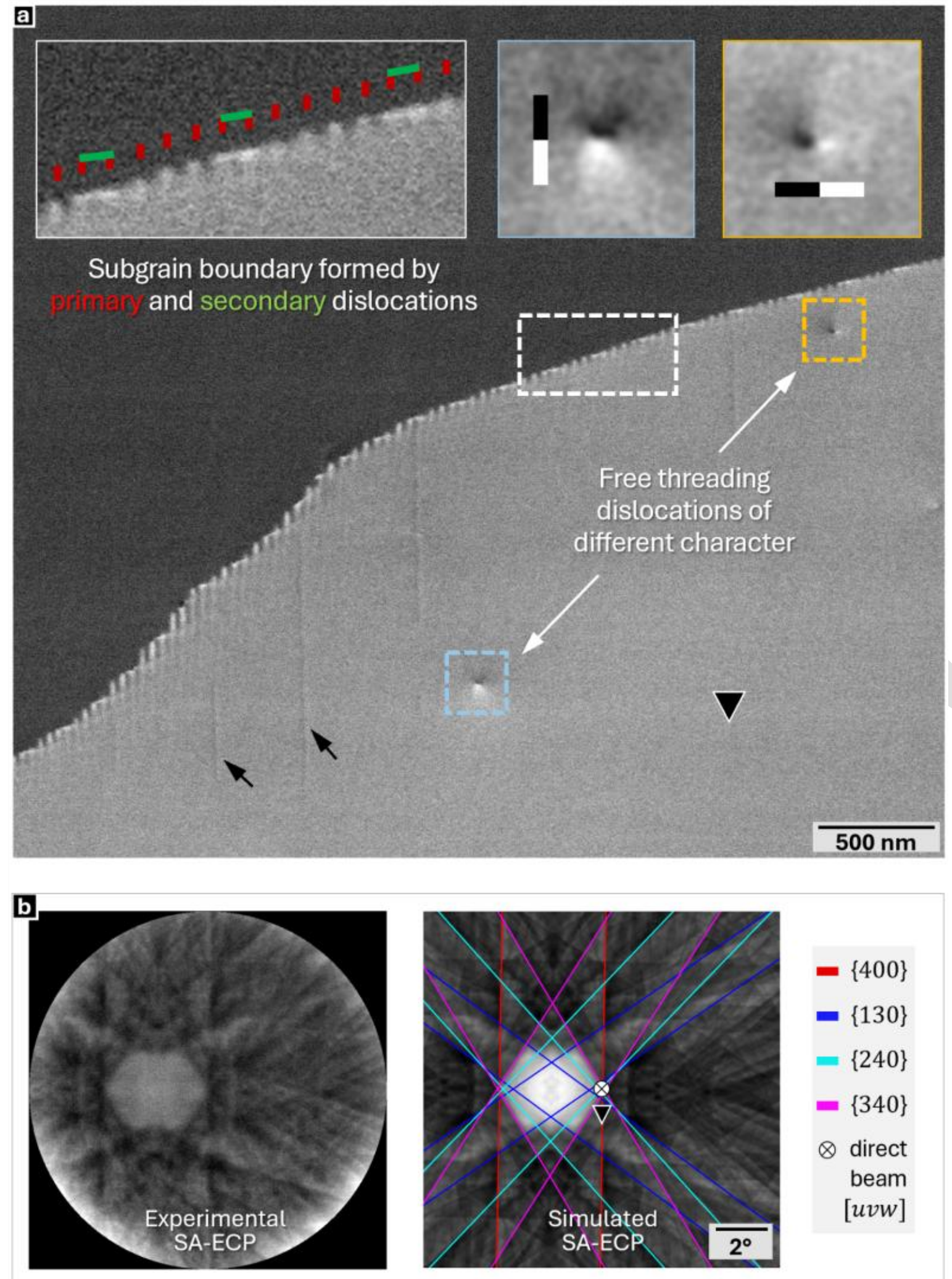

**Figure 9. (a)** ECCI micrograph showing a symmetrical subgrain boundary composed of primary (red) and secondary (green) dislocations. Two free dislocations with $\hat{l}$ = [001] are shown in the insets, having orthogonal black-to-white contrast directions relative to each other. The black arrows point to artifacts associated with carbon contamination from prior imaging. **(b)** Experimental and indexed/simulated SA-ECPs taken from the position marked with a triangle on the micrograph in (a).

# 4. Discussion

## 4.1. Characterizing Dislocations and Subgrain Boundaries Using SA-ECP Informed ECCI

Electron channeling pattern–informed ECCI, as implemented here, offers a practical route to survey defect architecture across representative bulk areas while using experimental and simulated patterns to navigate multiple ‘illumination’ conditions to reveal dislocations. Here, direct use of the indexed ECP has been employed to select the channeling condition and optimize contrast within the material. This approach is different from some other reported variants of ECCI, e.g., controlled ECCI (cECCI) (Zaefferer & Elhami, 2014) and orientation-optimized ECCI (ooECCI) (Miyajima et al., 2018), that use EBSD-informed geometries to place the beam in suitable diffraction conditions. ECP-informed ECCI has the advantage that the patterns are acquired in the same SEM geometry used for ECCI and therefore we can directly define the incident beam vector for the imaging condition. The smaller angular field of SA-ECPs also makes them especially useful for matching the specific band-edge condition that controls defect contrast, and combining this approach with complementary EBSD analysis helps in indexing zone axes.

Figure 10 provides a schematic overview of how dislocations of different character can produce contrast in ECCI in relation to the incoming beam direction [*uvw*] and the beam extinction distance, $\xi_g$:

(a) For dislocations with line directions near parallel to [*uvw*], their intersection with the surface produces a characteristic contrast against the background. Depending on the incident beam vector [*uvw*], character of the dislocations, and angle of observation, the dislocations may also appear as black-to-white twin lobe shapes (insets in Figures 8 and 9), comets (Figures 4, 5, and 6), line segments (Figure 8), or individual black/white 'pin-pricks' (Figure 4b).

(b) A surface-penetrating dislocation half loop shows a concentrated termination contrast where the loop meets the surface and a curved trace connecting the termination points due to the local strain field associated with the loop. If the half loop extends deeply into the material, then the dislocation loop will appear only as two dots.

(c) Subsurface loops with significant lengths of mixed character produce weaker, more diffuse contrast because their strain fields are partially screened by depth and by the mix of edge and screw components (Picard et al., 2012).

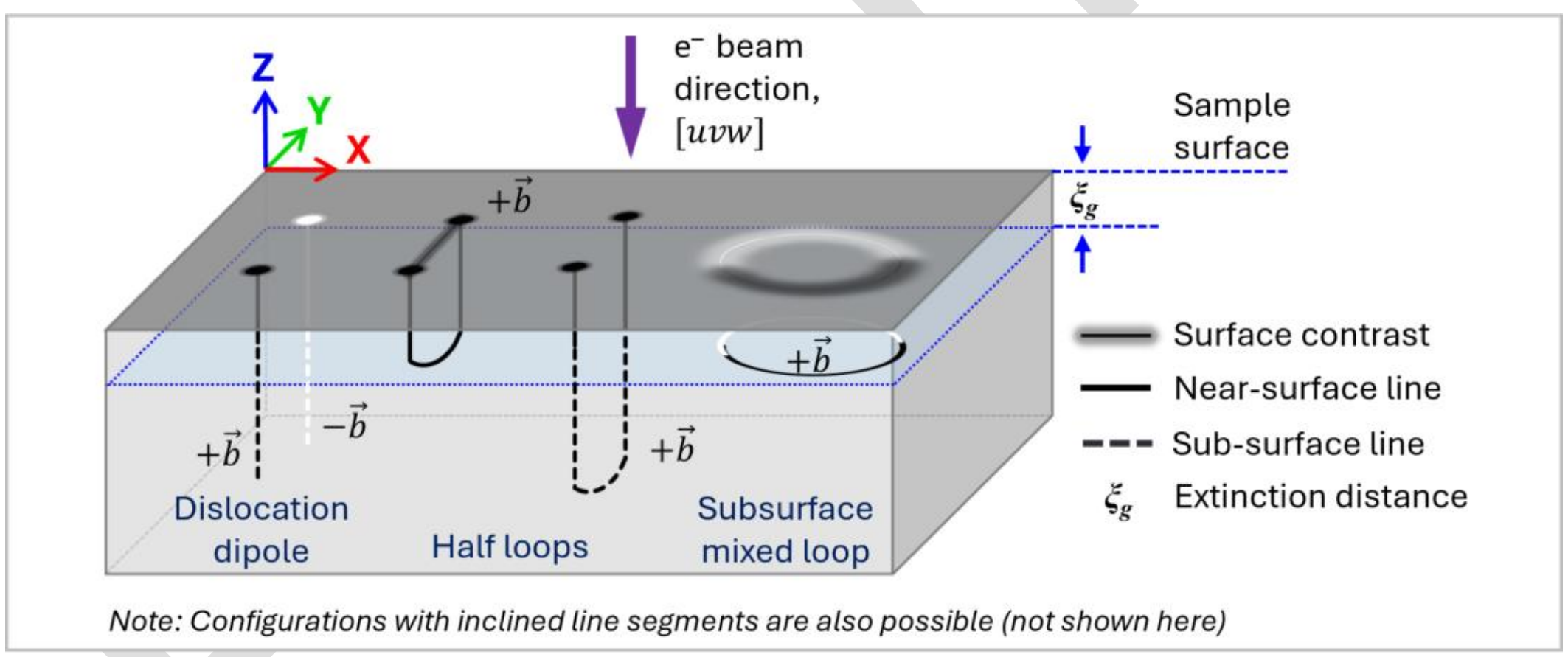

**Figure 10.** Simplified schematic depiction of how different crystallographic defects appear in ECCI with respect to the incoming electron ($e^-$) beam based on the beam extinction distance, $\xi_g$, and the incoming electron beam direction, [*uvw*].

Dislocations imaged by ECCI typically exhibit their strongest contrast where they intersect the free surface, with contrast diminishing progressively as the dislocation line extends into the bulk. This behavior reflects the finite interaction volume of backscattered electrons under channeling conditions and the rapid decay of channeling sensitivity with depth. Previous studies have shown that ECCI contrast in the SEM is generally limited to the upper ~100–200 nm of the specimen, depending on accelerating voltage and crystal orientation (Czernuszka et al., 1991; Simkin & Crimp, 1996) and that the crystallographic ECCI contrast is mostly dominated by electrons that are close to the primary beam energy (Qaiser et al., 2026). Consequently, only those segments of a dislocation that lie within this near-surface volume contribute appreciably to the observed contrast. The presence of elongate, comet-like features with visible line segments in Figure 4 therefore implies that the dislocations are inclined at shallow angles to the surface (Tunstall et al., 1964), such that a significant portion of their strain field remains within the effective channeling depth.

In Figure 6, ECCI contrast associated with the subgrain boundary has been explored using multiple imaging conditions at and near the [001] zone axis. For Figure 6a–d, the conditions have been selected to provide two 'pairs' of contrast. Figure 6a is on the intersection of the +{400}, +{240}, +{340}, and +{130} Kossel cone Kikuchi-band edges, and Figure 6b is the opposite edge pair on the opposite side of the [001] pole (a similar, yet different, contrast pair is provided for Figures 6c and 6d). The resulting contrast inversion in the ECCI micrographs highlights that the features observed are crystallographic, similar to reports by other authors (Kriaa et al., 2017; Kriaa et al., 2019; Spencer et al., 1972). Multiple illumination conditions, for the same area, can be used to distinguish crystallographic features from other types of feature, such as pores or amorphous material, as the contrast associated with a pore will not change with different channeling conditions. Further to these pair-based images, in Figure 6e, the dislocations are mostly invisible because we have targeted, in that case, the center of the [001] zone axis, which provides limited contrast variation around it (i.e., small changes in lattice orientation or strain, as associated with defects, would not change the contrast conditions much), and so for this [*uvw*], we are unable to see the strain gradient associated with most of the types of dislocations in that region.

As a note, an important practical limitation of ECP-informed ECCI is the spatial size of the 'selected area' in SA-ECP. In the SA-ECPs captured for this study, the estimated selected area is relatively large, likely due to spherical aberration and instrument beam optics. As a consequence, Figure 4d shows at least three distinguishable zone axes, which means that the acquired SA-ECP was collected from regions that contain an overlapping effect of the neighboring orientations in the crystal in an area on the order of 500 μm. As the SA-ECP has been acquired from a region straddling multiple subgrain boundaries, it exhibits multiple similar but displaced [001] zone axes. This displacement can be used to calculate the misorientation angle and axis, as previously noted by Joy et al. (1982). Smaller spatial size of the 'selected area' would allow collection of easier to index SA-ECPs from small grains/subgrains, as shown in prior studies (Guyon et al., 2015; Joy & Newbury, 1972; Kerns et al., 2020; van Essen et al., 1970).

## 4.2. Complex Dislocation Structures in Olivine

In general, low-angle grain boundaries have classically been divided into tilt, twist, or mixed character based on the angle, $\alpha$, between the misorientation axis and the grain-boundary normal. A twist boundary is an array of screw dislocations, where the boundary normal and misorientation axis are parallel. In contrast, a tilt boundary (a 'tilt wall') is an array of edge dislocations, and in this case the boundary normal and the misorientation axis are perpendicular. A boundary can also be of mixed character, consisting of a mixture of screw and edge dislocations, which may be a flat boundary segment where the misorientation axis is neither perpendicular nor parallel to the boundary normal, or the boundary can be curved, or the misorientation axis can vary along the boundary. For characterizing the boundary with two-dimensional EBSD or with ECPs taken from either side of the boundary, it is typically not possible to determine the full five parameters of the grain boundary, in the absence of assumptions about the grain-boundary habit plane (Lloyd et al., 2021) or a grain-boundary cross section. The full grain-boundary normal typically remains inaccessible and only the surface boundary trace (that is, the line of intersection between the boundary plane and the measurement surface) is visible. In the case shown in Figure 3, where the boundary trace is parallel to the misorientation axis, the boundary can be classified as a tilt wall. In a similar approach, in Figures 4b and 5a, the misorientation axis is along [010], which lies within the boundary plane, making the wall predominantly a tilt boundary. However, the spacing between adjacent dislocations is not uniform, and WBV results from the same region show multiple Burgers vectors and nonplanar segments, which overall makes it an 'asymmetric' tilt wall, as we indicate in Figure 5a. Here, 'asymmetric' means that the boundary plane normal is not arranged symmetrically about the misorientation axis and this configuration requires more than one dislocation set to accommodate the misorientation, though the misorientation axis still lies in the boundary plane. For the regions shown in Figures 8 and 9, the changing boundary plane evident in the images implies a mixed character of boundary that varies along the trace. The zig-zag geometry of the wall is consistent with an energy-minimizing response as discussed by Marquardt et al. (2015) to obtain as much area as possible of low-index surfaces, in this case, (100) planes. That zig-zag arrangement also echoes the observations of Lopez-Sanchez et al. (2021), who observed that subgrain rotation during dynamic recrystallization in olivine can produce low-angle tilt boundaries with rotation axes commonly parallel to [010] and composed of multiple edge dislocation systems. The non-planar geometries and complex structure of some subgrain boundaries in olivine is an important observation because the extent to which subgrain boundaries can act as dislocation sources or as obstacles to dislocation glide depends on the type and strength of dislocation interactions. For example, recent nanoindentation experiments found that a simple tilt subgrain boundary composed of a single set of edge dislocations with [001] Burgers vector was neither a potent source of dislocations nor a strong obstacle to dislocation glide (Avadanii et al., 2023). However, the more varied dislocation types present in more complex boundary structures offer increased potential for short-range interactions that may modulate rates of dislocation glide (Hansen et al., 2019, 2021; Wallis et al., 2017, 2020), such as the generation of pinning points to form dislocation sources and affect long-range interactions. These considerations motivate more extensive ECCI analysis of subgrain-boundary structures in olivine to establish the relative frequencies of their various characteristics.

The misorientation-axis analysis described above—especially when combined with complementary metrics, such as geometrically necessary dislocation (GND) density or WBV analysis—provides a means to assess whether a boundary geometry is consistent with the creep experiment loading geometry or instead reflects a pre-existing natural microstructure. In the case shown in Figure 3, the measured rotation axis for the boundary system is not consistent with the dislocation types expected to be generated under the direction of creep loading (applied along $[110]_c$), suggesting that the boundary was present prior to the creep experiment rather than formed by the imposed compression. This interpretation follows the same rationale commonly applied in geological studies that use misorientation-axis distributions to infer deformation history and active slip systems (e.g., Lloyd, 2004; Wallis et al., 2019). In brief, under $[110]_c$ compression, Schmid's law indicates that the (010)[100] slip system has the highest resolved shear stress (Schmid factor = 0.50), whereas the (001)[100] system has zero resolved shear stress in this geometry (Durham et al., 1977). Therefore, we would expect (010)[100] slip to dominate during deformation, not (001)[100]. In our case, however, the subgrain boundary has a [010] rotation axis (implying (001)[100] character), which is inconsistent with the loading direction. This mismatch indicates that the boundary was likely inherited from natural deformation rather than formed by the laboratory creep experiment.

The spatially resolved WBV analysis presented in Figure 7 supports the ECCI-based visual identification of several dislocation types and explains why the simple Read estimate could depart from the misorientations we measure by SA-ECP indexing and EBSD. In the case presented here, the net rotation arises from a superposition of different Burgers vectors, non-uniform spacing, and non-planar geometry. In contrast, Read's small-angle relation (and the related Read–Shockley description of subgrain-boundary energy) assumes an ideal planar wall composed of identical, equally spaced edge dislocations so that, for small tilts, $\theta \approx \frac{|\vec{b}|}{d}$ (Read & Shockley, 1950). In practice, low-angle boundaries often contain mixed dislocation characters and more than one Burgers vector, exhibit non-planar geometry, and have variable spacing and subsurface segments (Figures 4–7). These factors mean that a single Burgers vector, $\vec{b}$, and a single projected spacing, $d$, do not capture the full rotational content. This can be further understood by invoking the Frank–Bilby perspective and modern continuum descriptions of grain-boundary dislocation content. The Frank–Bilby framework links lattice rotation and a chosen boundary plane to the net interfacial Burgers-vector content that is required to accommodate that rotation, in effect predicting which combinations of dislocation types and line densities can reproduce the measured misorientation (Hirth et al., 2013). Continuum approaches extend this idea by treating the dislocation content as a distributed field, minimizing the energy of the equilibrium dislocations with respect to all Burgers vectors (Zhang et al., 2021). Long-range interactions, boundaries, and nearby defects further perturb the local strain field, so two-dimensional spacing measurements can be misleading if the true three-dimensional dislocation structure is complex.

Characterization of dislocation structures is necessary to ground-truth microphysics-based rheological models, which enable more reliable predictions of natural deformation than do traditional, phenomenological models. Long-range interactions among dislocations have been recognized to significantly influence the deformation behavior of olivine and therefore form the basis of recent microphysical models of dislocation-mediated deformation (e.g., Breithaupt et al., 2023; Hansen et al., 2019; Hansen et al., 2021). These long-range interactions depend on numerous microstructural variables that can be characterized using ECCI, which offers advantages over traditional techniques, such as being a non-destructive method. One key variable is (geometrically necessary) dislocation density, which is possible to quantify using EBSD analysis. With suitable assumptions, EBSD-based analyses can also be used to estimate the total dislocation density, including both GNDs and statistically stored dislocations (SSDs), through statistical treatment of the stress distribution and orientation gradients (Wilkinson et al., 2014). In some favorable materials, EBSD can even resolve individual dislocations, for example through virtual diode imaging (Hiller et al., 2023) or high-angular resolution EBSD methods (Ernould et al., 2022). However, as these microphysical models principally concern the density of mobile dislocations (Orowan, 1948), it is useful to be able to distinguish mobile and immobile dislocations. ECCI also allows dislocations of different types and signs of Burgers vector to be distinguished, which is important as different dislocation types exhibit different mobilities (Bai & Kohlstedt, 1992) and because dislocation dipoles may contribute to strain hardening (e.g., Mussi et al., 2017). The number of (suitable) dislocations, along with the separation of dislocations, also influences the rate of recovery (Breithaupt et al., 2023). The competition between recovery and storage processes is fundamental to recent models of olivine deformation (e.g.,

Breithaupt et al., 2023), and the activity of these processes may be better understood via observations of dislocation structures. Another critical factor to consider is dislocation morphology, as the rate of change of dislocation density with strain is inversely proportional to the width of a dislocation loop (Breithaupt et al., 2023) and the morphology of dislocation loops has been observed to depend on variables including crystallographic orientation, temperature, and deformation mechanism (e.g., Bai & Kohlstedt, 1992; Mussi et al., 2015; Mussi et al., 2017; Phakey et al., 1972). Therefore, ECCI-based analysis is extremely powerful, enabling constraints on many key variables, such as active deformation mechanisms, slip systems, storage and recovery processes, and the effect of deformation conditions and obstacles. As such, this tool has the potential to both validate existing microphysical models of deformation and contribute to the construction of models for understudied yet prevalent materials.

## 5. Conclusions

We demonstrate that ECP-based ECCI provides a minimally invasive route to map dislocation substructures in olivine, as an alternative (or complement) to TEM or oxidation decoration. Using multiple channeling conditions guided by pattern simulation and navigation software (AstroECP), we directly image and characterize subgrain boundaries, surface-threading dislocations, occasional closed loops, and localized dipole-like features in a single-crystal bulk sample of San Carlos olivine. The ECCI analysis presented is supported by EBSD data and WBV analysis, where the combined data reveal the presence of complex subgrain boundaries with multiple types of dislocations and deviation from Read's expression. These observations can link microstructural detail to macroscopic rheological behavior, particularly for informing microstructure-based models of dislocation-mediated deformation of Earth's upper mantle.

## Acknowledgments

The authors would like to thank Lars Hansen and David Kohlstedt of the Rock and Mineral Physics Laboratory, University of Minnesota, for providing the sample and the data associated with the deformation experiment, and would like to thank a range of funders that supported this collaborative work: Natural Sciences and Engineering Research Council of Canada (NSERC) [Discovery grant: RGPIN-2022-04762, 'Advances in Data Driven Quantitative Materials Characterization'] (TBB and HQ); the Canada Research Chair program ['Advancing Correlative Multimodal Electron Microscopy'] (TBB); UK Research and Innovation [Future Leaders Fellowship: MR/V021788/1, 'Microphysics of evolving rock viscosity in the seismic and glacial cycles'] (DW); and PEEF-2023 (HQ). Electron microscopy was performed within the Electron Microscopy Laboratory at the University of British Columbia, supported by funding from British Columbia Knowledge Fund (BCKDF) Canada Foundation for Innovation – Innovation Fund (CFI-IF) [#39798, 'AM+']. For EMsoft simulation of olivine, we acknowledge Lukas Berners and computing resources granted by RWTH Aachen University, Germany under project rwth-1308. This research was supported in part through the computational resources and services provided by Advanced Research Computing (ARC) at University of British Columbia (UBC), Canada.

## Data Availability Statement

Raw data used in the preparation of this manuscript are available via Zenodo repository (Qaiser et. al. 2026b).

## Conflict of Interest Disclosure

The authors declare that there are no conflicts of interest.